\documentclass{caosp305}

\usepackage{graphicx}
\usepackage{epstopdf}

\usepackage{natbib}
\articleNo{123}
\pubyear{2005}
\volume{35}
\volnumber{3}
\firstpage{1}
\received{January 16, 2018}

\def\BibTeX{{\rm B\kern-.05em{\sc i\kern-.025em b}\kern-.08em
             T\kern-.1667em\lower.7ex\hbox{E}\kern-.125emX}}

\begin{document}

%
\hauthor{A.\,David-Uraz {\it et al.}}

\title{Massive star winds interacting with magnetic fields on various scales}


%
\author{
        A.\,David-Uraz \inst{1}
      \and 
        V.\,Petit \inst{1}
      \and
        C.\,Erba \inst{1}
      \and
        A.\,Fullerton \inst{2}
      \and
        N.\,Walborn \inst{2}
      \and
        R.\,MacInnis \inst{1}
       }

%
\institute{
           Department of Physics \& Astronomy, University of Delaware\\
           Newark, DE 19716, USA, \email{adu@udel.edu}
         \and
           Space Telescope Science Institute\\
           Baltimore, MD 21218, USA
          }

\date{March 8, 2003}

\maketitle

\begin{abstract}
One of the defining processes which govern massive star evolution is their continuous mass loss via dense, supersonic line-driven winds. In the case of those OB stars which also host a surface magnetic field, the interaction between that field and the ionized outflow leads to complex circumstellar structures known as magnetospheres. In this contribution, we review recent developments in the field of massive star magnetospheres, including current efforts to characterize the largest magnetosphere surrounding an O star: that of NGC 1624-2. We also discuss the potential of the `analytic dynamical magnetosphere' (ADM) model to interpret multi-wavelength observations. Finally, we examine the possible effects of -- heretofore undetected -- small-scale magnetic fields on massive star winds and compare their hypothetical consequences to existing, unexplained observations.
\end{abstract}

%
\section{Introduction}
\label{intr}
One of the dominant processes driving massive star evolution is their continuous mass loss via stellar winds. These outflows are line-driven; despite having small abundances, metals possess numerous transitions (especially in the ultraviolet range of the electromagnetic spectrum) which can then lead to efficient driving in an expanding atmosphere since the associated Doppler shift counteracts the effects of saturation (for a concise review, see, e.g., \citealt{2011BSRSL..80...16O}). Thus, ultraviolet (UV) resonance lines constitute one of the most reliable observational diagnostics of massive star winds (e.g., \citealt{1994A&A...283..525P}). This wind-launching mechanism is well described using a power-law distribution of lines, leading to well-known scalings which allow us to derive relevant wind parameters \citep{1975ApJ...195..157C}; theoretical mass-loss rates and terminal velocities can easily be calculated for a given star given its stellar parameters \citep{2001A&A...369..574V}. 

However, the picture becomes a bit more complicated when we account for the effect of surface magnetic fields. About 7\% of OB stars possess detectable, globally organized, surface magnetic fields (e.g., \citealt{2015IAUS..307..342M,2017MNRAS.465.2432G}). These fields are found to be mostly dipolar and stable over large periods of time \citep{2014MNRAS.440..182S}.

\section{Magnetospheres}

Surface magnetic fields redirect and confine the stellar wind, as evidenced by a number of magnetohydrodynamic (MHD) simulations (e.g., \citealt{2008MNRAS.385...97U,2009MNRAS.392.1022U}), to form a circumstellar \textit{magnetosphere}. Material trapped in closed field loops is forced to co-rotate with the stellar surface, leading to various observational signatures.

\subsection{General structure and effects}

The structure of a magnetosphere is determined by the competition between the magnetic field and the wind momentum. 
Within closed field loops, wind material launched from both magnetic hemispheres accumulates around the magnetic equator, forming X-ray emitting shocks. Once cooled, it falls along the field lines back onto the surface. This corresponds to a \textit{dynamical magnetosphere}, or DM. However, around rapidly rotating stars, there is an added component as some material is centrifugally supported, preventing it from falling, thus forming dense co-rotating clouds, a so-called \textit{centrifugal magnetosphere}, or CM \citep{2008MNRAS.385...97U}.

While MHD simulations can provide a detailed description of magnetospheres, analytical models can help predict their behavior at a much smaller computational cost. For instance, the \textit{Rigidly Rotating Magnetosphere} model \citep{2005MNRAS.357..251T} provides a useful description of CMs and succesfully reproduces various observations. Likewise, the \textit{Analytic Dynamical Magnetosphere} model \citep{2016MNRAS.462.3830O} was developed to describe the time-averaged structure of DMs.

On top of confining and redirecting winds, magnetic fields also lead to mass loss quenching \citep{2008MNRAS.385...97U} and can brake surface rotation very efficiently \citep{2009MNRAS.392.1022U}, leading to important evolutionary consequences (see Keszthelyi et al., these proceedings).

\subsection{Observable consequences}

\begin{figure}
\centerline{\includegraphics[width=11.5cm,clip=]{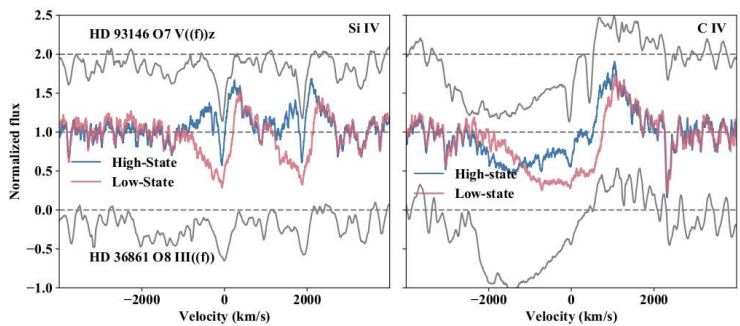}}
\caption{Comparison between the UV resonance lines of NGC 1624-2 (middle, at high and low state) and that of non-magnetic stars of similar spectral type. We can see that the magnetic star's line profiles show very different characteristics, and in particular, the C\textsc{iv} line is desaturated at high velocity, which might lead to an underestimation of the mass-loss rate using spherically symmetric wind models.}
\label{f1}
\end{figure}

According to the \textit{Oblique Rotator Model} \citep{1950MNRAS.110..395S}, the obliquity between the rotational and magnetic axes leads to periodic variations in the magnetospheric viewing angle. This can be seen for instance in H$\alpha$ profile variations, which can be modeled with the ADM model \citep{2016MNRAS.462.3830O}. This model can also explain the X-ray luminosity of magnetic massive stars due to magnetically confined wind shocks \citep{2014ApJS..215...10N}. Early attempts (Munoz et al., these proceedings) are being made to apply the ADM model to reproduce optical photometric variations, such as those seen in HD 191612 \citep{2011MNRAS.416.3160W}\footnote{Another proposed explanation for photometric variations in O stars involves wind blanketing and the latitudinal dependence of the mass flux (\citealt{2016A&A...594A..75K}, these proceedings)}. However, as mentioned previously, the most useful observational diagnostic to probe massive star winds is UV spectroscopy. The periodic variation of UV resonance line profiles has been detected in a number of magnetic O stars, and most recently in NGC 1624-2, the most strongly magnetized O-type star known to this day \citep{2012MNRAS.425.1278W}. Fig.~\ref{f1} shows two resonance lines from the UV spectra of NGC 1624-2 (obtained with \textit{HST/COS}) at high (nearly magnetic pole-on) and low (nearly magnetic equator-on) states. Their profiles show dramatic variations, as well as peculiarities that are not seen in non-magnetic stars.

While synthetic line profiles computed using spherically symmetric wind models have been compared to high state observations of magnetic O stars to yield wind parameters (e.g., \citealt{2013MNRAS.431.2253M}), ongoing efforts using the ADM model suggest that this technique leads to inaccurate mass-loss rate determinations.

\subsection{Complex fields}

The ADM formalism accounts for a large scale dipolar field, but can also be generalized for different magnetic topologies. Though rare, some massive stars exhibit complex magnetic fields, notably $\tau$ Sco \citep{2006MNRAS.370..629D}. The ADM model can be expanded to explain the observations of such stars (Fletcher et al., these proceedings).

\section{Small-scale magnetic fields}

While they have not yet been detected, small-scale magnetic fields (or magnetic spots) might arise as a consequence of the subsurface convection zone due to the iron opacity bump, and are expected to cause photospheric brightness variations \citep{2011A&A...534A.140C}. Such bright spots have been detected in $\xi$ Per \citep{2014MNRAS.441..910R} and $\zeta$ Pup \citep{2018MNRAS.473.5532R}. Hydrodynamical models \citep{1996ApJ...462..469C,2017MNRAS.470.3672D} show that bright spots might be the cause of the puzzling \textit{discrete absorption components} (DACs; e.g., \citealt{1996A&AS..116..257K}).

\section{Conclusions and future work}

Magnetic fields profoundly influence the density and velocity structure of massive star winds. This means that spherically symmetric wind models cannot lead to proper determinations of wind parameters. Future studies will test the ADM model and use it to determine the wind properties of magnetic massive stars.

While magnetic spots offer an attractive explanation of phenomena such as DACs, they have yet to be detected. Such an undertaking will require very deep magnetometry \citep{2013A&A...554A..93K} and requires a significant observational effort.

\acknowledgements
ADU gratefully acknowledges support from the \textit{Fonds qu\'{e}b\'{e}cois de la recherche en nature et technologies}. This research is based on observations made with the NASA/ESA {\it Hubble Space Telescope}, which is operated by the Association of Universities for Research in Astronomy, Inc., under NASA contract NAGS 5-26555.
Support for HST General Observer Program number GO-13734 was provided by NASA through a grant from the Space Telescope Science Institute, which is operated by the Association of Universities for Research in Astronomy, Incorporated, under NASA contract NAS5-26555.

\bibliography{demo_caosp305_adu}

\begin{thebibliography}{25}
\expandafter\ifx\csname natexlab\endcsname\relax\def\natexlab#1{#1}\fi

\bibitem[{{Cantiello} \& {Braithwaite}(2011)}]{2011A&A...534A.140C}
{Cantiello}, M. \& {Braithwaite}, J. 2011, {\it \aap}, {\bf 534}, A140, DOI
  10.1051/0004-6361/201117512

\bibitem[{{Castor} {et~al.}(1975){Castor}, {Abbott}, \&
  {Klein}}]{1975ApJ...195..157C}
{Castor}, J.~I., {Abbott}, D.~C., \& {Klein}, R.~I. 1975, {\it \apj}, {\bf
  195}, 157, DOI 10.1086/153315

\bibitem[{{Cranmer} \& {Owocki}(1996)}]{1996ApJ...462..469C}
{Cranmer}, S.~R. \& {Owocki}, S.~P. 1996, {\it \apj}, {\bf 462}, 469, DOI
  10.1086/177166

\bibitem[{{David-Uraz} {et~al.}(2017){David-Uraz}, {Owocki}, {Wade},
  {Sundqvist}, \& {Kee}}]{2017MNRAS.470.3672D}
{David-Uraz}, A., {Owocki}, S.~P., {Wade}, G.~A., {Sundqvist}, J.~O., \& {Kee},
  N.~D. 2017, {\it \mnras}, {\bf 470}, 3672, DOI 10.1093/mnras/stx1478

\bibitem[{{Donati} {et~al.}(2006){Donati}, {Howarth}, {Jardine}, {Petit},
  {Catala}, {Landstreet}, {Bouret}, {Alecian}, {Barnes}, {Forveille},
  {Paletou}, \& {Manset}}]{2006MNRAS.370..629D}
{Donati}, J.-F., {Howarth}, I.~D., {Jardine}, M.~M., {et~al.} 2006, {\it
  \mnras}, {\bf 370}, 629, DOI 10.1111/j.1365-2966.2006.10558.x

\bibitem[{{Grunhut} {et~al.}(2017){Grunhut}, {Wade}, {Neiner}, {Oksala},
  {Petit}, {Alecian}, {Bohlender}, {Bouret}, {Henrichs}, {Hussain},
  {Kochukhov}, \& {MiMeS Collaboration}}]{2017MNRAS.465.2432G}
{Grunhut}, J.~H., {Wade}, G.~A., {Neiner}, C., {et~al.} 2017, {\it \mnras},
  {\bf 465}, 2432, DOI 10.1093/mnras/stw2743

\bibitem[{{Kaper} {et~al.}(1996){Kaper}, {Henrichs}, {Nichols}, {Snoek},
  {Volten}, \& {Zwarthoed}}]{1996A&AS..116..257K}
{Kaper}, L., {Henrichs}, H.~F., {Nichols}, J.~S., {et~al.} 1996, {\it \aaps},
  {\bf 116}, 257

\bibitem[{{Kochukhov} \& {Sudnik}(2013)}]{2013A&A...554A..93K}
{Kochukhov}, O. \& {Sudnik}, N. 2013, {\it \aap}, {\bf 554}, A93, DOI
  10.1051/0004-6361/201321583

\bibitem[{{Krti{\v c}ka}(2016)}]{2016A&A...594A..75K}
{Krti{\v c}ka}, J. 2016, {\it \aap}, {\bf 594}, A75, DOI
  10.1051/0004-6361/201629222

\bibitem[{{Marcolino} {et~al.}(2013){Marcolino}, {Bouret}, {Sundqvist},
  {Walborn}, {Fullerton}, {Howarth}, {Wade}, \&
  {ud-Doula}}]{2013MNRAS.431.2253M}
{Marcolino}, W.~L.~F., {Bouret}, J.-C., {Sundqvist}, J.~O., {et~al.} 2013, {\it
  \mnras}, {\bf 431}, 2253, DOI 10.1093/mnras/stt323

\bibitem[{{Morel} {et~al.}(2015){Morel}, {Castro}, {Fossati}, {Hubrig},
  {Langer}, {Przybilla}, {Sch{\"o}ller}, {Carroll}, {Ilyin}, {Irrgang},
  {Oskinova}, {Schneider}, {D{\'{\i}}az}, {Briquet}, {Gonz{\'a}lez},
  {Kharchenko}, {Nieva}, {Scholz}, {de Koter}, {Hamann}, {Herrero},
  {Ma{\'{\i}}z Apell{\'a}niz}, {Sana}, {Arlt}, {Barb{\'a}}, {Dufton},
  {Kholtygin}, {Mathys}, {Piskunov}, {Reisenegger}, {Spruit}, \&
  {Yoon}}]{2015IAUS..307..342M}
{Morel}, T., {Castro}, N., {Fossati}, L., {et~al.} 2015, in IAU Symposium, Vol.
  307, New Windows on Massive Stars, ed. G.~{Meynet}, C.~{Georgy}, J.~{Groh},
  \& P.~{Stee}, 342--347

\bibitem[{{Naz{\'e}} {et~al.}(2014){Naz{\'e}}, {Petit}, {Rinbrand}, {Cohen},
  {Owocki}, {ud-Doula}, \& {Wade}}]{2014ApJS..215...10N}
{Naz{\'e}}, Y., {Petit}, V., {Rinbrand}, M., {et~al.} 2014, {\it \apjs}, {\bf
  215}, 10, DOI 10.1088/0067-0049/215/1/10

\bibitem[{{Owocki}(2011)}]{2011BSRSL..80...16O}
{Owocki}, S. 2011, {\it Bulletin de la Societe Royale des Sciences de Liege},
  {\bf 80}, 16

\bibitem[{{Owocki} {et~al.}(2016){Owocki}, {ud-Doula}, {Sundqvist}, {Petit},
  {Cohen}, \& {Townsend}}]{2016MNRAS.462.3830O}
{Owocki}, S.~P., {ud-Doula}, A., {Sundqvist}, J.~O., {et~al.} 2016, {\it
  \mnras}, {\bf 462}, 3830, DOI 10.1093/mnras/stw1894

\bibitem[{{Pauldrach} {et~al.}(1994){Pauldrach}, {Kudritzki}, {Puls}, {Butler},
  \& {Hunsinger}}]{1994A&A...283..525P}
{Pauldrach}, A.~W.~A., {Kudritzki}, R.~P., {Puls}, J., {Butler}, K., \&
  {Hunsinger}, J. 1994, {\it \aap}, {\bf 283}, 525

\bibitem[{{Ramiaramanantsoa} {et~al.}(2014){Ramiaramanantsoa}, {Moffat},
  {Chen{\'e}}, {Richardson}, {Henrichs}, {Desforges}, {Antoci}, {Rowe},
  {Matthews}, {Kuschnig}, {Weiss}, {Sasselov}, {Rucinski}, \&
  {Guenther}}]{2014MNRAS.441..910R}
{Ramiaramanantsoa}, T., {Moffat}, A.~F.~J., {Chen{\'e}}, A.-N., {et~al.} 2014,
  {\it \mnras}, {\bf 441}, 910, DOI 10.1093/mnras/stu619

\bibitem[{{Ramiaramanantsoa} {et~al.}(2018){Ramiaramanantsoa}, {Moffat},
  {Harmon}, {Ignace}, {St-Louis}, {Vanbeveren}, {Shenar}, {Pablo},
  {Richardson}, {Howarth}, {Stevens}, {Piaulet}, {St-Jean}, {Eversberg},
  {Pigulski}, {Popowicz}, {Kuschnig}, {Zoc{\l}o{\'n}ska}, {Buysschaert},
  {Handler}, {Weiss}, {Wade}, {Rucinski}, {Zwintz}, {Luckas}, {Heathcote},
  {Cacella}, {Powles}, {Locke}, {Bohlsen}, {Chen{\'e}}, {Miszalski}, {Waldron},
  {Kotze}, {Kotze}, \& {B{\"o}hm}}]{2018MNRAS.473.5532R}
{Ramiaramanantsoa}, T., {Moffat}, A.~F.~J., {Harmon}, R., {et~al.} 2018, {\it
  \mnras}, {\bf 473}, 5532, DOI 10.1093/mnras/stx2671

\bibitem[{{Silvester} {et~al.}(2014){Silvester}, {Kochukhov}, \&
  {Wade}}]{2014MNRAS.440..182S}
{Silvester}, J., {Kochukhov}, O., \& {Wade}, G.~A. 2014, {\it \mnras}, {\bf
  440}, 182, DOI 10.1093/mnras/stu306

\bibitem[{{Stibbs}(1950)}]{1950MNRAS.110..395S}
{Stibbs}, D.~W.~N. 1950, {\it \mnras}, {\bf 110}, 395, DOI
  10.1093/mnras/110.4.395

\bibitem[{{Townsend} \& {Owocki}(2005)}]{2005MNRAS.357..251T}
{Townsend}, R.~H.~D. \& {Owocki}, S.~P. 2005, {\it \mnras}, {\bf 357}, 251, DOI
  10.1111/j.1365-2966.2005.08642.x

\bibitem[{{Ud-Doula} {et~al.}(2008){Ud-Doula}, {Owocki}, \&
  {Townsend}}]{2008MNRAS.385...97U}
{Ud-Doula}, A., {Owocki}, S.~P., \& {Townsend}, R.~H.~D. 2008, {\it \mnras},
  {\bf 385}, 97, DOI 10.1111/j.1365-2966.2008.12840.x

\bibitem[{{Ud-Doula} {et~al.}(2009){Ud-Doula}, {Owocki}, \&
  {Townsend}}]{2009MNRAS.392.1022U}
{Ud-Doula}, A., {Owocki}, S.~P., \& {Townsend}, R.~H.~D. 2009, {\it \mnras},
  {\bf 392}, 1022, DOI 10.1111/j.1365-2966.2008.14134.x

\bibitem[{{Vink} {et~al.}(2001){Vink}, {de Koter}, \&
  {Lamers}}]{2001A&A...369..574V}
{Vink}, J.~S., {de Koter}, A., \& {Lamers}, H.~J.~G.~L.~M. 2001, {\it \aap},
  {\bf 369}, 574, DOI 10.1051/0004-6361:20010127

\bibitem[{{Wade} {et~al.}(2011){Wade}, {Howarth}, {Townsend}, {Grunhut},
  {Shultz}, {Bouret}, {Fullerton}, {Marcolino}, {Martins}, {Naz{\'e}}, {Ud
  Doula}, {Walborn}, \& {Donati}}]{2011MNRAS.416.3160W}
{Wade}, G.~A., {Howarth}, I.~D., {Townsend}, R.~H.~D., {et~al.} 2011, {\it
  \mnras}, {\bf 416}, 3160, DOI 10.1111/j.1365-2966.2011.19265.x

\bibitem[{{Wade} {et~al.}(2012){Wade}, {Ma{\'{\i}}z Apell{\'a}niz}, {Martins},
  {Petit}, {Grunhut}, {Walborn}, {Barb{\'a}}, {Gagn{\'e}},
  {Garc{\'{\i}}a-Melendo}, {Jose}, {Moffat}, {Naz{\'e}}, {Neiner}, {Pellerin},
  {Penad{\'e}s Ordaz}, {Shultz}, {Sim{\'o}n-D{\'{\i}}az}, \&
  {Sota}}]{2012MNRAS.425.1278W}
{Wade}, G.~A., {Ma{\'{\i}}z Apell{\'a}niz}, J., {Martins}, F., {et~al.} 2012,
  {\it \mnras}, {\bf 425}, 1278, DOI 10.1111/j.1365-2966.2012.21523.x

\end{thebibliography}
\end{document}